\documentclass[referee]{aa}

\usepackage{txfonts}

\usepackage{graphicx}

\begin{document}

\def\P{\bar{\Phi}}

\def\st{\sigma_{\rm T}}

\def\vk{v_{\rm K}}

\def\sles{\lower2pt\hbox{$\buildrel {\scriptstyle <}
   \over {\scriptstyle\sim}$}}

\def\sgreat{\lower2pt\hbox{$\buildrel {\scriptstyle >}
   \over {\scriptstyle\sim}$}}






\title{The role of reconnection in the pulsar magnetosphere}

\author{Ioannis Contopoulos}
\institute{Research Center for Astronomy, Academy of Athens, 
GR-11527 Athens, Greece, 
\email{icontop@academyofathens.gr}}

\date{Received 28 June 2006 / Accepted 2 October 2006}

\abstract{

The present work is our first attempt to understand the 
role of reconnection in the pulsar magnetosphere.
Our discussion is based on the observationaly
inferred fact that, as the pulsar spins down, 
the region of closed corotating dipolar field lines grows
with time. This implies that reconnection must
take place in the magnetosphere. We argue
that non-dissipative reconnection along the equatorial
current sheet allows for the continuous channeling
of pulsar spindown energy into particle energy,
all the way from the light cylinder to the
pulsar wind termination shock, and we
propose that this effect may
account for the low $\sigma_{shock}$ values inferred
by observations.
We present a simple model that allows us
to relate the magnetic diffusivity in the
equatorial current sheet to an observable
pulsar parameter, the braking index $n$.
When $n\sim 1$, the global
structure of the magnetosphere approaches that
of a relativistic split monopole where the
pulsar spindown energy is carried by the
electromagnetic field. However,
for values of $n\ \sgreat\ 1.5$, almost all field lines 
close inside the pulsar wind termination shock,
and thus most of the electromagnetic pulsar
spindown energy flux is effectively transformed 
into particle energy in the equatorial current sheet.

}

\maketitle

\section{Introduction}

In the past few years, several steps have been taken that improved
our understanding of the electrodynamic structure of the pulsar 
magnetosphere (\cite{M91}; \cite{CKF99} 
(hereafter CKF); \cite{M99}; \cite{B99}; \cite{S06})).
A canonical paradigm begins to emerge, namely that
a magnetized rotating neutron star with polar magnetic field
$B_*\sim 10^{13}$~G, mass $M_*\sim 1.4M_\odot$, radius $r_*\sim 10$~km, and
angular velocity $\Omega$ loses rotational kinetic energy
at a rate
\begin{equation}
\dot{E}_{kin}=\frac{2}{5}M_* r_*^2 \Omega\dot{\Omega}
\label{Ekin}
\end{equation}
through electromagnetic torques in its magnetosphere. 
The poloidal structure of the 
pulsar magnetosphere remains dipolar-like (closed) from 
the surface of the rotating central
neutron star to a distance 
\begin{equation}
r_{open}\leq r_{lc}=\frac{c}{\Omega}
\end{equation}
($r_{lc}$ is the so called `light cylinder').
Beyond $r_{open}$, the poloidal structure of the
magnetosphere becomes monopole-like (open). The total amount of
open poloidal magnetic flux is of the order of
\begin{equation}
\Psi_{open}=1.23\frac{\pi B_* r_*^3}{r_{open}}\ .
\label{Psiopen}
\end{equation}
The magnetic field along the open monopole-like 
magnetic field lines is
wound backwards in the azimuthal direction by the neutron star
rotation, and a toroidal magnetic field is generated. 
The source of the toroidal field is a poloidal
electric current whose distribution along the open field lines 
is given by 
\begin{equation}
I(\Psi)\approx 
-\frac{\Omega\Psi}{4\pi}\left(2-\frac{\Psi}{\Psi_{open}}\right)\ .
\label{IPsi}
\end{equation}
when $0\leq \Psi\leq\Psi_{open}$, and zero outside.
This current distribution implies the existence
of a singular return electric current sheet flowing along
the equator and along the separatrix between open and
closed field lines. Beyond distances greater
than a few times the light cylinder distance, the dominant
field component becomes the azimuthal one.
Finally, the electromagnetic spindown luminosity is given by
\begin{equation}
L= \frac{\Omega^2 \Psi_{open}^2}{6\pi^2 c}\ .
\label{L}
\end{equation}
Note that eqs.~\ref{Psiopen}, \ref{IPsi}
and \ref{L} are valid for an aligned rotator. The case of
an oblique rotator remains under investigation (\cite{B99};
\cite{S06}; \cite{C06}).
We have introduced here  a cylindrical system of coordinates 
$(r,z,\phi)$ aligned with the central rotating neutron star.
We have also introduced the magnetic flux $\Psi$ as
\begin{equation}
{\bf B}_p\equiv \nabla\times\left(\frac{\Psi}{2\pi r}\hat{\phi}\right)\ ,
\end{equation}
where ${\bf B}_p$ is the poloidal component of the magnetic field.
Furthermore, $\dot{(\ldots)}\equiv \partial(\ldots)/\partial t$.

In developing the above paradigm, there has been 
almost general agreement within the astrophysics community
that the global structure of the pulsar magnetosphere may be described
by the ideal MHD formalism, except for regions of finite extent 
such as regions of particle acceleration along the magnetic field 
(polar gaps (e.g. \cite{DH82}), slot gaps (\cite{HM05}), 
outer gaps (e.g. \cite{RY95}),
light cylinder dissipation regions (e.g. \cite{MS94}), etc.), and
current sheets which, as we will see, seem to be
of fundamental importance to the electrodynamics of the system.
The above regions where perfect coupling between particles and 
the magnetic field breaks down influence the solution of the
ideal MHD problem through the boundary conditions. 
We have dealt with the presence of particle acceleration gaps
in previous papers (\cite{C05}; \cite{CS06}).
We have shown there that particle acceleration
along the magnetic field `consumes' a certain
amount of electric potential $V_{gap}\sim 10^{13}$~V 
(e.g. \cite{HA01}) from 
the total `available' electromotive potential
\begin{equation}
V=\frac{\Psi_{open}}{2\pi r_{lc}}\sim 10^{13-17}\mbox{Volt}
\label{V}
\end{equation}
that develops on the surface of the rotating neutron star,
and therefore, only the remaining amount is 
available to establish the global ideal MHD magnetosphere.
In general, $V_{gap}\ll V$,
except for pulsars near their death line where $V_{gap}\sim V$,
in which case, the presence of particle acceleration gaps has important
implications for the neutron star spindown (see \cite{CS06}).

In the present work, we deal with the second region where
ideal MHD breaks down,
namely the global magnetospheric current sheet. 
In order to understand how reconnection may work in the
pulsar magnetosphere, we begin in \S~2 with the presentation
of a simple magnetostatic analog.
In \S~3, we argue that reconnection does take place, and in \S~4
we show that reconnection extends from the light cylinder
to the pulsar wind termination shock,
and that it modifies the global
structure of the pulsar magnetosphere. In \S~5 we
discuss the significance of reconnection for the acceleration 
of the pulsar wind to the low $\sigma_{shock}$ values inferred by
observations. Finally, in \S~6 we present a summary of our
conclusions.

\section{A simple magnetostatic analog}

The role of current sheets in ideal MHD is to
help accomodate the requirements imposed on the system by the 
specific boundary conditions of the particular problem
under investigation. 
In order to illustrate our point, let us consider a simple
non-relativistic axisymmetric system consisting of a magnetostatic
dipole surrounded by an equatorial thin 
disk of half-thickness 
\begin{equation}
h\ll r\ ,
\end{equation}
with a hole inside a radius $r=r_{open}$. 
The magnetic field structure may be obtained as a solution
of the force-free (and current-free outside the disk)
problem described by 
\begin{equation}
\nabla\times {\bf B} = 0\ ,
\label{1}
\end{equation}
or equivalently
\begin{equation}
\frac{\partial^2\Psi}{\partial r^2}-\frac{1}{r}
\frac{\partial\Psi}{\partial r}+
\frac{\partial^2\Psi}{\partial z^2}=0\ ,
\label{pulsareq1}
\end{equation}
with boundary conditions a) dipolar field at the center, 
and b) horizontal field right above and below the thin disk, 
\begin{equation}
B_r(r< r_{open}; z=0)=0\ ,\ \mbox{and}
\label{bc1}
\end{equation}
\begin{equation}
B_z(r\geq r_{open}; z=h)=B_z(r\geq r_{open}; z=-h)=0\ .
\label{bc2}
\end{equation}
These particular boundary conditions determine the
magnetic field structure shown in fig.~1, which further implies the
presence of a current sheet in the azimuthal direction all along the
surface of the infinitely thin disk.
Note that, because of the symmetry of the problem, 
$B_r(r;z)=-B_r(r;-z)$, $B_\phi(r;z)=-B_\phi(r;-z)$, 
and $B_z(r;z)=B_z(r;-z)$.

The reader may have already noticed that the field structure 
obtained in the infinite conductivity case, fig.~1, is
{\em very similar} to the poloidal magnetic field structure of
the CKF solution\footnote
{The equatorial boundary conditions, eqs.~\ref{bc1} \& \ref{bc2},
are the same as in the CKF problem 
(described by eq.~\ref{pulsareq} below).
Therefore a solution of eq.~\ref{pulsareq1} `kills' the first part
in the l.h.s. of eq.~\ref{pulsareq}. 
Moreover, around and beyond the light cylinder,
the r.h.s. balances the second part in the l.h.s. of eq.~\ref{pulsareq}.}.
Moreover, one realizes that the assumption
of infinite conductivity (i.e. no field reconnection) has
been central in obtaining the CKF solution, where we required
that field lines open up beyond a certain radius 
$r_{open}$, and never again reconnect.

Now, let us assume that, once the solution shown in fig.~1 
is established,
we somehow introduce a small amount of resistivity 
(or equivalently diffusivity) in the disk. 
Formerly open magnetic field lines will now close accross the
disk and will slowly begin to diffuse inward. If the diffusion is slow
enough, we may approximate the magnetic field evolution 
as a sequence of force-free equilibria 
outside the resistive thin disk,
coupled to the magnetic field threading the disk.
The evolution of the latter is determined by the
induction equation inside the disk,
\begin{equation}
\dot{\bf B}=
-c\nabla\times {\bf E}=
\nabla\times ({\bf v}\times {\bf B}-\eta \nabla\times {\bf B})\ .
\label{induction}
\end{equation}
We have made use here of the generalized Ohm's law
\begin{equation}
{\bf E}+\frac{{\bf v}\times {\bf B}}{c}=
\frac{\eta}{c}\nabla\times {\bf B}\ ,
\end{equation}
where ${\bf v}$ and $\eta$ are the matter velocity and magnetic
diffusivity in the disk respectively. 
The $z$-component of eq.~\ref{induction} yields
\begin{equation}
\dot{B}_z(z=0)=\frac{1}{r}\frac{\partial}{\partial r}
\left(-rv_r B_z-\eta r \frac{\partial B_r}{\partial z}\right)\ ,
\label{Bdot}
\end{equation}
or equivalently
\begin{equation}
\dot{\Psi}
=-2\pi rv_r B_z -2\pi\eta r\frac{\partial B_r}{\partial z}
\approx -2\pi rv_r B_z -2\pi\eta r\frac{B_r(z=h)}{h}
\label{2}
\end{equation}
(we have made use here of the fact that the disk is thin,
and hence $\partial/\partial z \gg \partial/\partial r$).
When a flow field is present in the disk, 
as is the case for the pulsar magnetosphere, 
the magnetic flux evolution will be determined
by the imbalance between advection and diffusion
(first and second terms in the r.h.s. of eq.~\ref{2}
respectively). In the present
section, we will only consider the simplest case ${\bf v}=0$. 
A radial equatorial flow with ${\bf v}\approx c\hat{\bf r}$
will be introduced in the next section.

One may evolve the above system of equations
(eqs.~\ref{pulsareq1} \& \ref{2}), as follows:
\begin{enumerate}
\item Start with any initial distribution 
$B_z(r\geq r_{open};z=0;t=0)$ in the disk
(in the example discussed here, we take $B_z(r\geq r_{open};z=0;t=0)=0$).
\item Solve eq.~\ref{pulsareq1} 
to obtain the field structure above and below
the disk. This also yields the distribution 
$B_r(r\geq r_{open};z=h;t=0)$ on the surface of the disk.
\item Obtain the distribution $B_z(r\geq r_{open};z=0;t=t+dt)$ by evolving
eq.~\ref{2} for one time step ${\rm d}t$.
\item Repeat steps 1 to 3 to obtain the field evolution 
at all times $t$.
\end{enumerate}
A sequence of field equilibria is shown in fig.~2, where
timescales are normalized to the characteristic
time $r_{open}h/\eta$. One sees that, if we allow for
an infinitesimally small amount of magnetic diffusivity in the disk, 
given enough time, the field structure will approach asymptotically
the structure of an isolated magnetostatic dipole
(in that case, $B_r(z=h)\approx B_r(z=0)=0$, and thus, $\dot{\Psi}=0$).
We emphasize that this would not have been possible, had we
insisted that the disk were infinitely conductive, as is the
case in fig.~1.

Up to now, we have ignored
the large radial magnetic stresses that develop inside the thin
disk in the presence of reconnection (see fig.~2).
The infinite conductivity solution shown in fig.~1 is stress-free,
and therefore it is valid even for a non-rigid disk.
However, in that case the solution would be unstable, and
any small perturbation would result in global magnetospheric
oscillations. Any small amount of dissipation would
damp those oscillations, and the field
structure would approach asymptotically the structure
of the isolated magnetostatic dipole, which is the structure
with the lowest possible magnetic field energy
(asymptotically, $B\propto r^{-3}$ for the dipole,
and $\propto r^{-2}$ for the solution shown in fig.~1 respectively).
We conclude that the sequence of field equilibria shown in
fig.~2 will take place only if the equatorial disk is
`rigid enough' to hold the radial magnetic stress that
develops because of the presence of reconnection.

\section{An argument for reconnection}

We will now argue that reconnection does indeed take
place in the pulsar magnetosphere.
We would like to note
that, although magnetic reconnection is believed to play
a very important role in various systems of physical and
astrophysical interest (flares in the solar corona,
planetary magnetospheres, laboratory plasmas, etc.),
we know very little about the physical processes that
control reconnection in other astrophysical systems
like pulsar magnetospheres and accretion disks. 
Although it is customary
to assume conditions of infinite conductivity,
in the case of the pulsar magnetosphere, we will see that
this is not justified by observations.

One thing we are certain about pulsars is that they spin down.
We also believe that the spindown is due to magnetospheric torques.
Equating the magnetospheric energy loss (eq.~\ref{L})
to the observed stellar spindown
(eq.~\ref{Ekin}), one obtains a relation between $\dot{\Omega}$
and $\Omega$, namely that
\begin{equation}
\dot{\Omega}=
\frac{5}{8}\frac{B_*^2 r_*^4}{M_* c}\frac{\Omega}{r_{open}^2}
\label{Omegadot}
\end{equation}
Furthermore, 
in the case of the six pulsars of table~1, $\dot{\Omega}$ is
found to be proportional to some power of $\Omega$
\begin{equation}
\dot{\Omega}\propto \Omega^n\ ,
\label{braking}
\end{equation}
where $n$, the braking index, has been observed to have values
between 1 and 3. This implies that
\begin{equation}
\frac{r_{open}}{r_{lc}}=\left(\frac{\Omega}{\Omega_o}\right)
^{\frac{3-n}{2}}< 1
\label{ropen}
\end{equation}
(and in fact, in some cases, $r_{open}/r_{lc}\ll 1$).
We have assumed here that $n$ remains unchanged
throughout the pulsar's lifetime (a simplification),
and that $r_{open}=r_{lc}$ at pulsar birth when
$\Omega=\Omega_o$.
If we assune that there is no reconnection,
we imply that $\Psi_{open}$, and consequently $r_{open}$,
remain fixed to the value they attained
at pulsar birth. This is equivalent to $\dot{\Omega}\propto \Omega$, 
and therefore $n=1$, which is not what is observed.

We conclude that in order to account for braking index values
$n>1$ in our standard picture of electromagnetic pulsar spindown, 
reconnection has to take place somewhere in the pulsar magnetosphere
outside $r_{open}$.
As we will see in the next section, 
the natural place to look for it is the
equatorial current sheet that develops outside $r_{open}$.
We may obtain an estimate of reconnection, by
calculating the rate of decrease of open magnetic flux 
as the pulsar spins down,
\begin{equation}
\dot{\Psi}_{open}=
\frac{(n-1)}{2}\Psi_{open}\frac{\dot{\Omega}}{\Omega}
\sim -10^{-3}\frac{\Psi_{open}}{\mbox{year}}
\label{rec}
\end{equation}
(see table~1). We would like to emphasize here that the
fact that $\dot{\Psi}_{open}\neq 0$ has nothing to do
with the fact that the pulsar spins down. Even in the
theoretical case where some artificial source of energy
kept $\Omega$ unchanged, 
as long as $r_{open}< r_{lc}$ the closed line region would
grow against the open line region at a rate determined
by eq.~\ref{2}. This realization allows us
to relate the magnetic diffusivity in the equatorial
current sheet immediately outside $r_{open}$
to the observable pulsar parameters, namely that
\begin{equation}
\frac{\eta}{hc}\sim
\left.
\frac{|\dot{\Psi}|}{2\pi rr_{lc} \Omega
B_r(r;z=h)}\right|_{r=r_{open}^+}
+\left. \frac{|B_z|}{B_r(r;z=h)}
\right|_{r=r_{open}^+}
\ .
\label{etah}
\end{equation}
We have introduced here a radial equatorial flow with 
$v_r\approx c$ beyond $r=r_{open}$. This accounts for the
high Lorentz factor electron-positron outflow that
is expected to flow along the open magnetic field lines.
Inside the corrotating closed line region $B_r(z=0)=0$ 
(see CKF), and therefore 
just outside $r_{open}$, $B_r(z=h)$ is also very
small (figure~11 in \cite{T06}). 
Moreover, in the ideal MHD case (CKF), $B_z(z=0)$ 
is taken equal to zero beyond $r_{open}$. In the presence
of reconnection we expect that $|B_z|\ll |B_r|$
in the equatorial region beyond $r_{open}$ 
(see solution in the next section), therefore,
we may ignore the second term in the r.h.s. of eq.~\ref{etah}.
We thus propose a simple model where $\eta/hc$ depends only on
the braking index $n$ through the following approximate
relation,
\begin{equation}
\frac{\eta}{hc}=\kappa (n-1)\ ,
\label{etahscaling}
\end{equation}
where the parameter $\kappa$ will be determined in \S~5.
This model is consistent with our understanding that,
the further $n$ differs from unity, the
greater the role of reconnection in the magnetosphere.

We would like to remind the reader that ``a braking index that
reflects the physics of the pulsar braking mechanism
can only be measured for the very youngest pulsars 
which show either little glitch activity or have been observed 
for many years, in which case the statistical effect of glitches 
can be determined'' (\cite{H04}). This is
why we have restricted our discussion to the case
of the six pulsars of table~1. However, in view of the recent
measurements of $\ddot{\Omega}$ for many more pulsars
(e.g. \cite{JG99}, \cite{H04}), one might be tempted
to also consider values of $n$ very different from the ones 
in table~1. In the context of our present model, values of
$n\gg 1$ may be interpreted as a closed line region growing
through very efficient (anomalous?) reconnection
$\sim n$ times faster than the dipole canonical value\footnote{
Other physical mechanisms such as magnetic field
decay (e.g. \cite{UV93}) may also be responsible for
higher than normal values of $n$.}.
On the other hand, values of $n\ll 1$ may be obtained when
a strong stellar wind acts to open up formerly closed magnetic
field lines (e.g. \cite{HCK99}). In this latter
picture, reconnection is irrelevant. One may even speculate that
the wide range of $n$ values may reflect a long
period (100-1000 years) cyclic evolution of the system
between the above two extremes (see \cite{BBK06}).

\section{Continuous equatorial reconnection}

In the previous section, 
we have only considered reconnection in the
poloidal magnetic field component, which is generated by a
corresponding toroidal electric current.
The MHD rotator, however, develops also a
toroidal magnetic field component $B_\phi$ with a corresponding
poloidal electric current (eq.~\ref{IPsi}) such that
\begin{equation}
B_\phi=\frac{2I(\Psi)}{cr}\ .
\label{Bphi}
\end{equation}
The most natural place one would expect reconnection
to take place is the global magnetospheric current sheet
where the electric current density is the greatest.
We remind the reader
that an important element of the force-free ideal
MHD solution is the presence
of an equatorial current sheet 
(poloidal + toroidal)
that extends from the surface
of the star, along the separatrix between open and closed field lines,
and along the equator. The poloidal component of this current sheet 
constitutes the return current for the
globally distributed (i.e. non-singular) electric current
that flows along the open field lines from the stellar surface
to infinity. This particular electric current distribution 
and therefore the corresponding
return current sheet was obtained as an eigenvalue of the
problem by requiring that there is no magnetic field discontinuity 
at the mathematical singularity of the light cylinder.
We would like to emphasize this point again: given the
particular equatorial boundary condition that we have chosen
in CKF, namely that $B_z(r>r_{open};z=0)=0$, the existence of the
above global current sheet is inevitable.

As we will now see, the latter choice of equatorial boundary condition
is not 100\% valid when we introduce nonzero
magnetic diffusivity $\eta$ and half thickness $h$ in the
equatorial poloidal current sheet.
Because of the symmetry of the problem,
\begin{equation}
{\bf B}(r;z=0)=B_z\hat{\bf z}\ ,
\end{equation}
and therefore, when $B_z(z=0)\neq 0$ a radial component of the
magnetospheric poloidal electric field 
${\bf E}\equiv -r\Omega\hat{\bf \phi}\times {\bf B}_p/c$
develops along the surface of the equatorial current sheet, namely
\begin{equation}
E_r(r;z=h)=-\frac{\Omega r}{c}B_z(r;z=h)
\approx -\frac{\Omega r}{c}B_z(r;z=0)\ .
\label{Er}
\end{equation}
For a slowly evolving magnetosphere, the $\phi$-component
of the induction equation for the equatorial current sheet
(eq.~\ref{induction}) implies that
\begin{equation}
\left. \frac{\partial E_r}{\partial z}\right|_{z\sim 0}
\approx 0\ ,
\end{equation}
which further implies that $E_r(z=h)\approx E_r(z=0)$,
or equivalently
\begin{equation}
\Omega r B_z \approx 
\eta\frac{\partial B_\phi}{\partial z}
\approx \frac{\eta B_\phi(z=h)}{h}
=\frac{\eta}{hc}\frac{2 I(\Psi)}{r}\ .
\label{slip}
\end{equation}
This may be rewritten as
\begin{equation}
\frac{\partial\Psi}{\partial r} \approx
\frac{\eta}{hc}\frac{4\pi I(\Psi)}{\Omega r}
\label{Psir}
\end{equation}
for $z=0$ and $r\geq r_{open}$. Note that interior to $r_{open}$,
$B_r(r;z=0)=0$, and therefore the field structure is close
to dipolar with
\begin{equation}
\frac{\partial\Psi}{\partial r} \approx
-\frac{\Psi}{r}\ .
\end{equation}
$I(\Psi)$ is determined similarly to CKF as an eigenvalue of 
the solution of the force-free magnetosphere outside the 
current sheet (see below). 

The physical meaning of eq.~\ref{slip} is that
the vertical ($z$) component of the electromagnetic (Poynting)
energy flux is nonzero at the surface of the
equatorial current sheet,
\begin{equation}
\left. \frac{c}{4\pi}{\bf E}\times {\bf B}\right|_z
=\frac{c}{4\pi}E_r B_\phi(z=h)\ ,
\end{equation}
and hence electromagnetic energy is continuously channeled to the
equatorial current sheet where it is transformed into some
different (non-electromagnetic) form. As we will see in the
next section, it is transformed into particle kinetic energy.
Another way to look at eq.~\ref{slip} is that
field lines `corotate' with the neutron star at angular velocity
$\Omega$, and therefore, particles `trapped' along field lines
would corotate at speeds $r\Omega$ (and at distances $r>r_{lc}$
these would exceed the speed of light) 
{\em unless} the equatorial region is
diffusive enough for the field line to `slip through' the
particles at a rate given by the r.h.s. of eq.~\ref{slip}.
This is a generalization
of the argument that led us in CKF to propose the boundary
condition that $B_z(r>r_{lc};z=0)=0$. 

If one knows the distribution of $\eta/h$ with distance
along the equatorial current sheet, 
eq.~\ref{Psir} together with the
special relativistic generalization of eq.~\ref{1},
\begin{equation}
(\nabla\times {\bf B})\times {\bf B} +4\pi \rho_e {\bf E}= 0\ ,
\label{4}
\end{equation}
or equivalently
\begin{equation}
\left(1-\frac{r^2}{r_{lc}^2}\right)
\left(\frac{\partial^2\Psi}{\partial r^2}-\frac{1}{r}
\frac{\partial\Psi}{\partial r}+
\frac{\partial^2\Psi}{\partial z^2}\right)
+\frac{2}{r}\frac{\partial\Psi}{\partial r}=-
\frac{4I(\Psi)}{c^2}
\frac{{\rm d}I(\Psi)}{{\rm d}\Psi}\ ,
\label{pulsareq}
\end{equation}
yield the structure of the pulsar magnetosphere
all the way to the wind termination shock distance
$r_{shock} \sim 10^9 r_{open}$.
We may expect that, 
the presence of equatorial reconnection modifies
the CKF solution only slightly, with magnetic flux
gradually reconnecting along the equator.

In order to obtain a solution for the structure
of the global pulsar magnetosphere, we will make
the simplification that $\eta/h$ remains constant
at all distances, equal 
to the value given by eq.~\ref{etahscaling}.
We argued that, for small enough values of $\eta/h$,
the structure of the magnetosphere will not differ
much from that of the CKF solution, and in
particular, the poloidal electric current distribution
will not differ much from eq.~\ref{IPsi}.
In that case, eq.~\ref{Psir} yields
\begin{equation}
\Psi=\Psi_{open}
\frac{2}{1+\left(\frac{r}{r_{open}}\right)^{2\kappa (n-1)}}
\label{Psir2}
\end{equation}
for $z=0$ and $r\geq r_{open}$. 
Given the equatorial distribution of $\Psi$,
we obtain the global magnetospheric structure shown in
figs.~3 \& 4 for 2 different values of $n$.
Note that we have assumed here that, since
reconnection is proportional to the electric current
density, the perfect particle-field coupling
required in deriving eq.~\ref{pulsareq} remains
valid everywhere throughout the magnetosphere, {\em except}
in regions with very large electric current density,
as is the case for the equatorial current sheet.

We are thus in a position
to estimate the amount of magnetic flux that will close 
inside the pulsar wind termination shock as a function of 
the diffusivity parameter $\eta/hc$, and more specifically,
as a function of the difference $(n-1)$.
As we will see in the next section,
this result may be related to the
acceleration mechanism in the pulsar wind.

\section{Equatorial reconnection and wind acceleration}

We have shown that the small amount of reconnection implied
by the observations of pulsar spindown requires the presence
of a radial electric field component along the
equatorial return current sheet in the pulsar magnetosphere
(eq.~\ref{Er}). In other words, a potential drop
\begin{equation}
V_{rec} = \left(1-\frac{\Psi(r_{shock})}{\Psi_{open}}\right) V
\label{Vrec}
\end{equation}
develops along the equatorial current sheet, from about
the light cylinder to the distance of the pulsar wind termination
shock. Here, $\Psi(r_{shock})$
is the amount of magnetic flux that does not close
accross the equatorial current sheet by the time the
pulsar wind reaches the termination shock.

According to reconnection theory (e.g. \cite{PF00}),
when the value of the reconnection electric field
(in our case $E_r$ as given in eq.~\ref{Er}) is greater
than a characteristic value, the Dreicer field defined as
\begin{equation}
E_{D}=
\frac{2\pi e^3 n_e \ln\Lambda}{kT_e}\ ,
\end{equation}
particles that enter the current sheet get
accelerated by the electric field and do not
interact during their acceleration process.
Here, $n_e$ and $T_e$ are the particle number density and
temperature in the current sheet respectively;
$\ln\Lambda$ is the Coulomb logarithm.
For number densities up to a few orders of
magnitude greater than the local Goldreich-Julian
number density (\cite{GJ69}), the radial electric field
inside the pulsar magnetosphere equatorial current sheet
is found to be orders of magnitude larger
than the Dreicer field, as is the case
in the geomagnetic tail (see \cite{LW84} for a review).
In this situation, the orbits of individual particles
must be calculated by explicit integration of their
equations of motion, and the acceleration mechanism 
is essentially a finite gyroradius effect (i.e. non-MHD).
According to that scenario, energy in the magnetic field
lines that reconnect accross the current sheet may
be transformed into the kinetic energy of the
accelerated particles without dissipation into other
forms of energy (\cite{S87} refers to this process as 
`inertial dissipation'). This result is quite encouraging.
The equatorial potential drop (eq.~\ref{Vrec}) 
is available to accelerate some particles (electrons/positrons)
up to Lorentz factors
\begin{equation}
\Gamma_{shock} = \frac{e V}{m_e c^2}
\left(1-\frac{\Psi(r_{shock})}{\Psi_{open}}\right)\sim
\left(1-\frac{\Psi(r_{shock})}{\Psi_{open}}\right)
 10^{11}\ .
\label{Lorentz}
\end{equation}
We have obtained here a perfect particle accelerator,
all the way from the light cylinder to the
termination shock. It seems thus natural to associate
the particle dominated wind that is inferred observationaly
in the case of the Crab pulsar 
(\cite{RG74}; \cite{KC84}) with the particles that 
belong to the equatorial current sheet
(mainly positrons/electrons for an aligned/counter-aligned 
rotator respectively). In a realistic pulsar,
the finite angle between the magnetic and rotation axes,
and the finite scale height $h$ of the equatorial current sheet
will lead to a finite opening of the equatorial particle
dominated pulsar wind. 
Note that our picture of gradual poloidal + toroidal magnetic
field reconnection is different from the 
reconnection of the toroidal magnetic field
that has been proposed as a source of the pulsar wind 
acceleration in a `stripped wind' (\cite{C90}; \cite{M94};
\cite{LK01}).

We have started to investigate the relativistic dissipationless
acceleration of particles in the equatorial current sheet
(Contopoulos, Gontikakis \& Eftymiopoulos, in preparation).
This by itself is a very promising new area of research.
The dynamics of the equatorial current sheet will also be studied.
In particular, in analogy to the issue of the rigidity of the
conducting disk in our simple magnetostatic
analog of \S~2, we need to address the issue of the
stability of the current sheet in the presence of the
radial magnetic stresses that develop accross it due to
reconnection. It is interesting that the
ideal MHD CKF solution implies that the current sheet
is electricaly charged (the vertical component of the electric
field on the surface of the current sheet, 
$E_z=\Omega r B_r/c$, changes sign accross it), and therefore,
we expect that the radial electric field $E_r$ that
develops along it contributes to
counteract the radial magnetic stress.
Note that dissipationless particle acceleration
due to reconnection in 
the equatorial current sheet was first discussed
in \cite{RCL05}. Our presentation, however, differs from 
theirs in that we tried to relate the toroidal magnetic field
reconnection to the global redistribution of the poloidal
magnetic field in the pulsar magnetosphere. This may
also be seen from the fact that our equatorial radial
electric field scales as $(r/r_{lc})^{-1-\epsilon}$ 
with $\epsilon\ll 1$, whereas in their work it drops
much faster with distance as $(r/r_{lc})^{-2}$.

We will conclude here with a discussion
of the overall energy balance in the pulsar magnetosphere.
In our simple picture of constant $\eta/h$, 
we may estimate the fraction of the total spindown energy
flux that remains in the from of electromagnetic field 
energy by the time we reach the pulsar wind termination
shock,
\[
\frac{L(\Psi(r_{shock}))}{L(\Psi_{open})}\approx 
\frac{\int_{\Psi=0}^{\Psi(r_{shock})}I(\Psi){\rm d}\Psi}
{\int_{\Psi=0}^{\Psi_{open}}I(\Psi){\rm d}\Psi}
= \frac{3}{2}\left(\frac{\Psi(r_{shock})}{\Psi_{open}}\right)^2
\left[1-\frac{1}{3}\frac{\Psi(r_{shock})}{\Psi_{open}}\right]
\]
\begin{equation}
\sim 2\left[1+3\left(\frac{r_{shock}}{r_{open}}\right)
^{2\kappa (n-1)}\right]
\cdot \left[1+\left(\frac{r_{shock}}{r_{open}}\right)
^{2\kappa (n-1)}\right]^{-3}
\label{Lratio}
\end{equation}
(e.g. \cite{G05}). 
The rest of the spindown energy flux
is gradually channeled to the particles along the
equatorial sheet. It is very tempting to
associate the inferred values of $\sigma_{shock}$
(defined as the ratio
of energy flux in the electromagnetic field to the energy flux
in the particles at the pulsar wind termination shock), with the
of energy flux estimated above remaining in the form of
electromagnetic field energy to the energy flux
transferred to the particles in the equatorial current sheet.
It is interesting that if we fix our only free parameter
to the value 
\begin{equation}
\kappa=0.05\ ,
\end{equation}
we are able to produce estimates for both the
Crab and Vela observationaly inferred values of $\sigma_{shock}$
(see table~1).

\section{Summary and conclusions}

We have presented here a new picture for the global structure
of the pulsar magnetosphere, where field lines remain closed
even beyond the light cylinder (figs.~3 \& 4). 
This is only possible if we take into account
the effect of magnetic field reconnection accross the
equatorial magnetospheric current sheet. We argue that
observations of braking index values $n> 1$ 
allow us to infer that
reconnection is indeed at work in the pulsar magnetosphere,
and we produce an estimate of the reconnection
time scale (eq.~\ref{rec}; table~1).
In fact, we propose to quantify the amount of effective
magnetic diffusivity as being proportional to the difference
$(n-1)$ (eqs.~\ref{etah}, \ref{etahscaling}). 
In our model, the neutron star spindown energy
carried in the form of electromagnetic Poynting flux
is gradually directed towards the equator where it is
channeled to the particles along the equatorial sheet with
no radiation losses (eq.~\ref{Lratio}). 
Without considering the details
of dissipationless particle acceleration, we argue
that individual particles may reach very high Lorentz factors
(eq.~\ref{Lorentz}). Finally, we discuss the issue of the
very low $\sigma_{shock}$ values inferred at the pulsar 
wind termination shock. 
In our picture the regions of field and particle
dominated flow are physically separated, and
we can estimate the ratio of the respective energy fluxes.
If then we assume that these regions somehow mix at the
termination shock, we are able to estimate the
values of $\sigma_{shock}$.
When we calibrate our diffusivity parameter 
(eq.~\ref{etahscaling}) in order to obtain a $\sigma_{shock}$
value relevant to the Crab pulsar wind termination shock,
all $\sigma_{shock}$ values lie in the range
$10^{-2}-10^{-3}$ {\em except} for the Vela pulsar
where $\sigma_{shock}$ is believed to have a value
close to unity. Thus, the Vela pulsar is singular, because it has
a much higher value of $\sigma_{shock}$, together
with a much lower value of the braking index $n$.
Our present work offers a physical picture where the above
two observed parameters are related. It is interesting that
Vela is also singular in that it shows a very strong glitch
activity. This may contribute to a more effective
loading of the magnetosphere with charges, and thus explain
the reduced effective magnetic diffusivity
in the equatorial current sheet.

We conclude that collisionless relativistic reconnection
along the equatorial current sheet may play a fundamental
role in determining both the near and large scale structure
of the pulsar magnetosphere, and in accounting for the
acceleration of the pulsar wind, therefore, it certainly
deserves further investigation.

\acknowledgements{We would like to thank Roger Romani and
Anatoly Spitkovsky for their strong criticism that led us 
to reconsider certain parts of the present work. 
We would also like to thank Christos Efthymiopoulos and 
Constantinos Gontikakis for introducing us to the field 
of collisionless reconnection.}

\newpage

\begin{table}
\begin{tabular}{lccccccc}
 & n & $\frac{\eta}{hc}$ & $r_{open}$ &
$r_{lc}$ & $\frac{\Psi_{open}}{|\dot{\Psi}_{open}|}$ 
& $\sigma_{shock}$ & $\frac{L_{field}}{L_{particles}}$ \\ 
&  & & ($10^8$~cm) & ($10^8$~cm) & ($10^3$~years)  & &\\ \hline
Crab & 2.515 & 0.08 & 1.2 & 1.6 & 3 & $10^{-2}-10^{-3}$ 
& $8\times 10^{-3}$  \\
Vela & 1.4 & 0.02 & 0.7 & 4.3 & 113 & $1-10^{-1}$ 
& $7\times 10^{-1}$\\
PSR 1509-58 & 2.84 & 0.09 & 5.8 & 7.2 & 3 & & $4\times 10^{-3}$ \\
PSR B0540-69 & 2.140 & 0.06 & 1.2 & 2.4 & 6 & & $4\times 10^{-2}$ \\
PSR J1846-0258 & 2.65 & 0.08 & 8.4 & 15.5 & 2 & & $8\times 10^{-3}$ \\
PSR J1119-6127 & 2.91 & 0.10 & 16.5 & 19.5 & 3 & & $4\times 10^{-3}$ \\
\hline
\end{tabular}
\caption{Characteristic parameters for the 6 pulsars with measured values
of the braking index $n$. 
In estimating $r_{open}$ we assumed an initial period of 1~msec at
pulsar birth. We took here $\eta/hc=0.05(n-1)$, and
$r_{shock}\sim 0.1$~pc.}
\end{table}

\newpage

\begin{figure}
\includegraphics[angle=270,scale=1.]{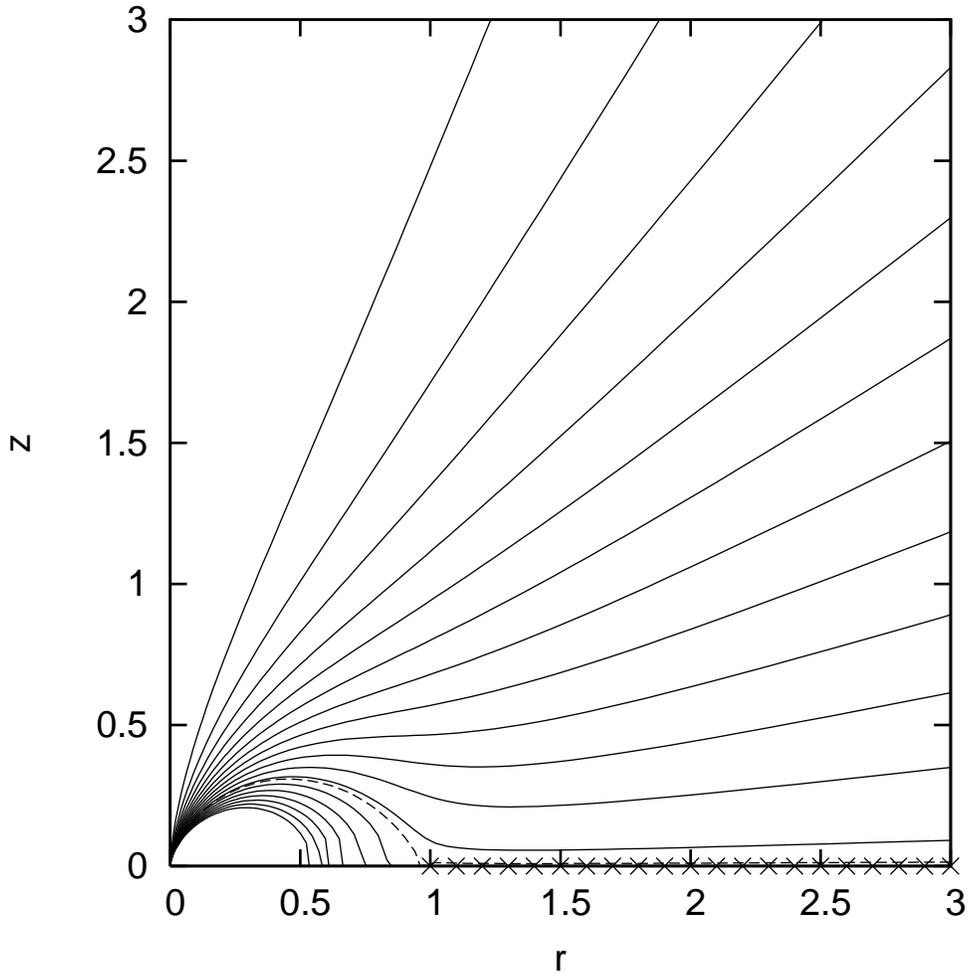}
\caption{The poloidal magnetic field structure of
a magnetic dipole surrounded by an infinitely conducting
thin disk with a hole inside some radius $r_{open}$.
The disk is denoted schematically with a line of x's
along the equator. Length units are normalized to $r_{open}$.
Solid lines correspond to magnetic flux intervals of
$0.1\Psi_{dipole}$, where $\Psi_{dipole}$ is defined
as the amount of magnetic flux of a pure dipole that would
cross the equator at distances greater or equal to $r_{open}$.
The dashed line denotes the separatrix $\Psi=\Psi_{open}=
1.23\Psi_{dipole}$ between open and closed field lines.
There is a close similarity between the present solution and the
poloidal structure of the CKF magnetosphere (CKF).}
\label{fig1}
\end{figure}

\begin{figure}
\includegraphics[angle=270,scale=0.5]{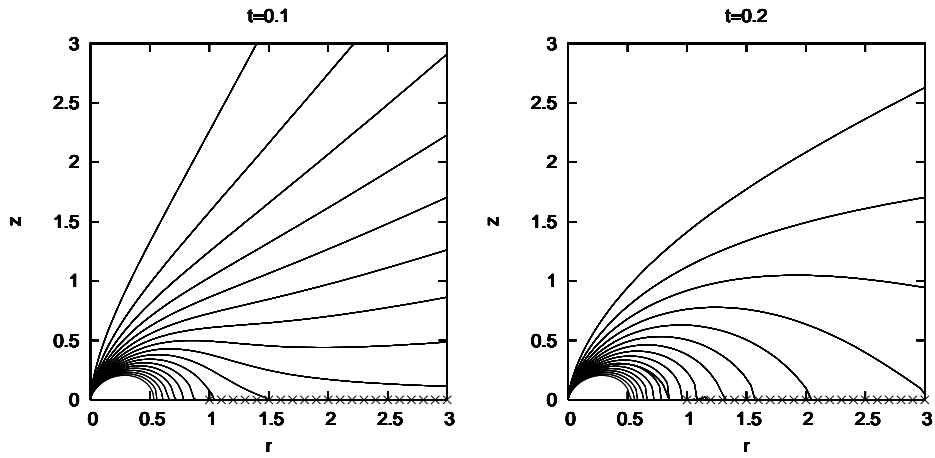}
\caption{A typical sequence of poloidal magnetic field
configurations when we introduce magnetic diffusivity in
the equatorial disk of fig.~1. Units same as in fig.~1.
Time is normalized
to the characteristic timescale
$r_{open}h/\eta$, where $\eta$ and $h\ll r$ are the disk
magnetic diffusivity and half thickness respectively.
Note the development of large radial stresses in
the interior of the rigid diffusive disk. 
In the limit $t\rightarrow \infty$,
the magnetic field configuration 
approaches that of a pure magnetostatic dipole, which
is a stress-free configuration.}
\label{fig2}
\end{figure}

\begin{figure}
\includegraphics[angle=270,scale=1.]{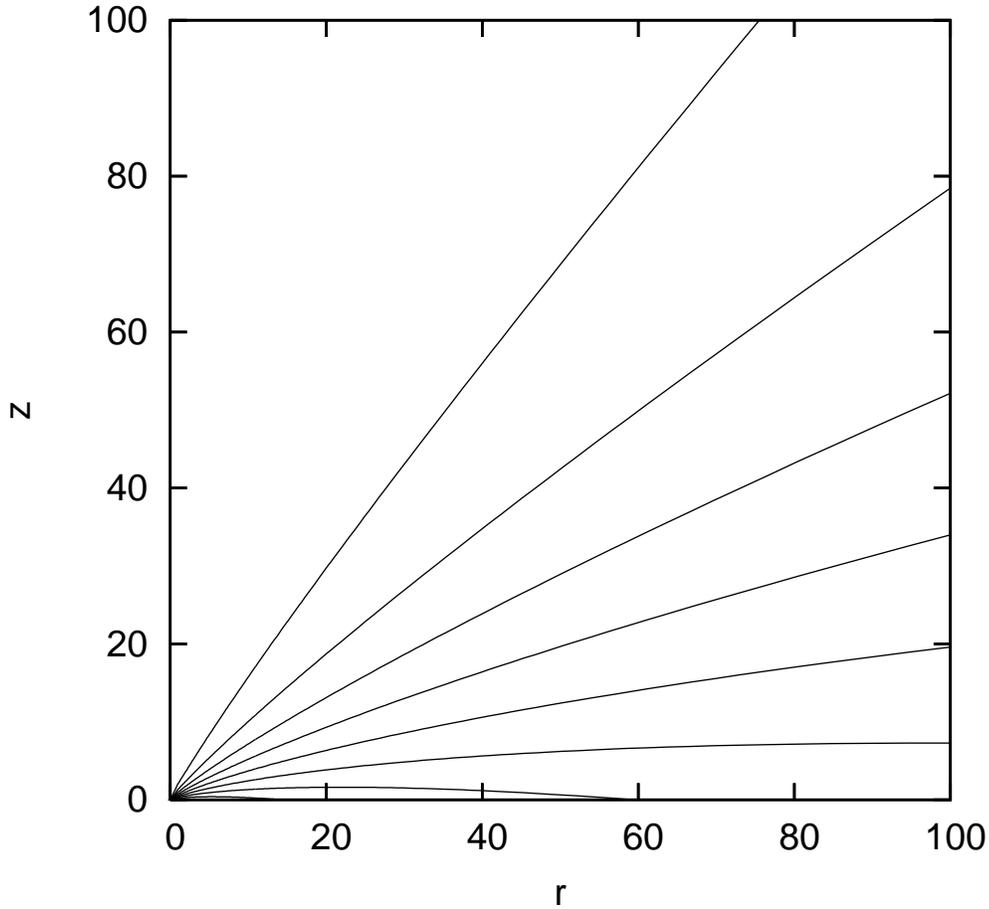}
\caption{The modified structure of the large scale axisymmetric pulsar
magnetosphere when we account for magnetic diffusivity
in the equatorial current sheet.
Units same as in fig.~1.
We set $\eta/h=0.08c$, a value that we believe is relevant
for the Crab pulsar magnetosphere (see text).
One sees the closing of the magnetosphere
accross the equatorial current sheet.
Note that an amount of poloidal magnetic flux equal to
about $0.3\Psi_{open}$ reconnects accross the
equatorial current sheet inside about $100 r_{lc}$.}
\label{fig3}
\end{figure}

\begin{figure}
\includegraphics[angle=270,scale=1.]{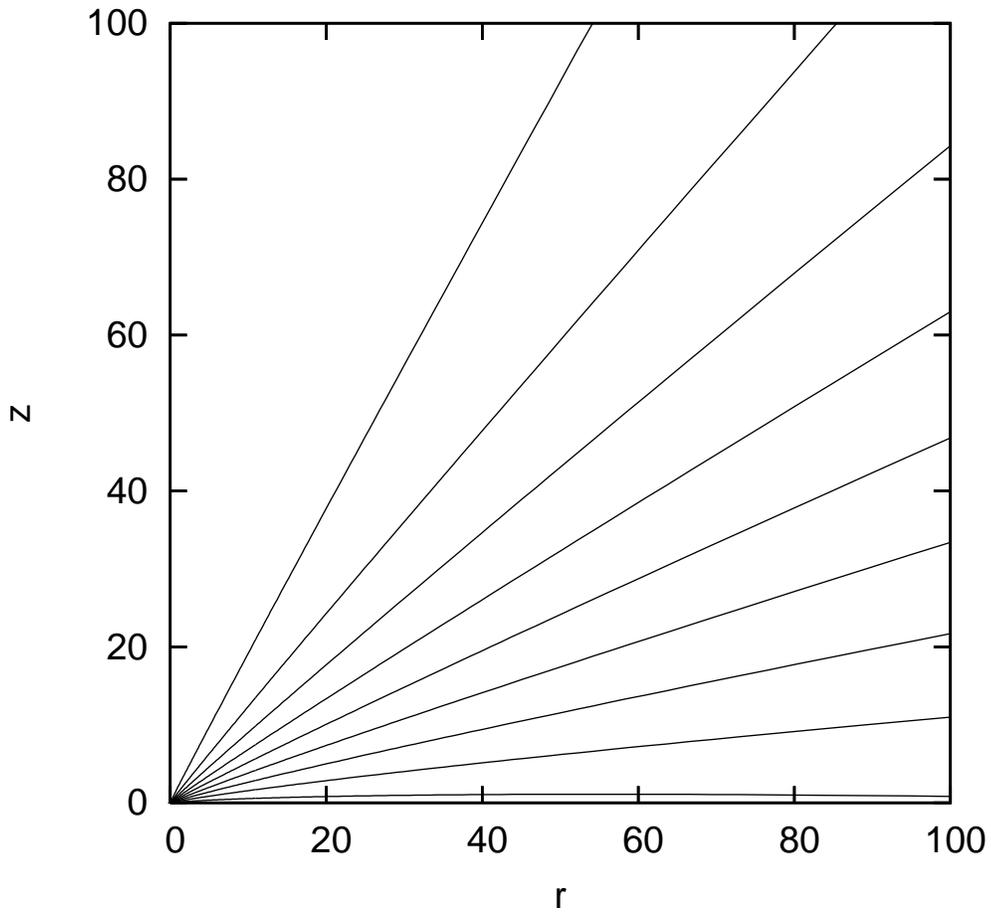}
\caption{Same as fig.~3 with $\eta/h=0.02c$, 
a value that we believe is relevant
for the Vela pulsar magnetosphere (see text).
The closing of the magnetosphere
accross the equatorial current sheet is slower 
when the current sheet magnetic diffusivity is smaller.
Note that the amount of poloidal magnetic flux 
that reconnects inside about $100 r_{lc}$
is equal to about $0.1\Psi_{open}$, i.e. smaller than in fig.~3}
\label{fig4}
\end{figure}

\end{document}